\documentclass{iopart}
\usepackage{graphicx}

\begin{document}
\title{Dynamics of a Bose-Einstein Condensate in an Anharmonic Trap}
\author{H Ott, J Fort\'agh and C Zimmermann}

\address{Physikalisches Institut der Universit\"at T\"ubingen\\
Auf der Morgenstelle 14, 72076 T\"ubingen, Germany}

\begin{abstract}
We present a theoretical model to describe the dynamics of
Bose-Einstein condensates in anharmonic trapping potentials. To
first approximation the center-of-mass motion is separated from
the internal condensate dynamics and the problem is reduced to the
well known scaling solutions for the Thomas-Fermi radii. We
discuss the validity of this approach and analyze the model for an
anharmonic waveguide geometry which was recently realized in an
experiment \cite{Ott2002c}.

\end{abstract}
%\pacs{03.75.Fi, 03.75.Be, 34.50.Dy, 75.70.-i}

It is well known, that for a cloud of particles that is moving in
a parabolic potential the center-of-mass motion is completely
decoupled from the internal excitations
\cite{Kohn1961a,Dobson1994a}. This holds in the presence of
interaction between the particles and therefore also applies to
Bose-Einstein condensates. Most of the experimental work with
Bose-Einstein condensates has been done in parabolic potentials.
The static properties of a condensate, see e.\,g.
Ref.\,\cite{Mewes1996a,Ernst1998a}, as well as the dynamical
properties \cite{Jin1996a,Mewes1996b} have been investigated and
were found to be in good agreement with the theoretical
predictions
\cite{Stringari1996a,Fetter1996c,Ruprecht1996a,Castin1996a}.
Higher order contributions to the trapping potential were either
avoided or turned out to be negligible due to the large scale
magnetic field generating elements. Anharmonic configurations were
theoretically considered for a Gaussian \cite{Martikainen2001a}
and a box-like \cite{Villain2001a} potential and were
experimentally realized in the case of a magnetic mirror
\cite{Arnold2002a}. Current progress in generating Bose-Einstein
condensates in magnetic microtraps \cite{Ott2001a,Hansel2001b} in
which the trapping potential in general differs from the trivial
parabolic shape raises the questions how the dynamics of the
trapped condensate are affected by the anharmonicity of the
potential. In a recent experiment, the dynamics of a Bose-Einstein
condensate in an anharmonic waveguide were studied in detail
\cite{Ott2002c} and the theory in this paper has been developed to
describe the experimental data.

In anharmonic traps the center-of-mass motion of a condensate is
coupled to the internal dynamics. In the rest frame of the
condensate, the local potential curvature changes during the
center-of-mass oscillation and collective modes of the condensate
are excited. We present a theoretical model which allows us to
describe the experimental data to high accuracy. The model is
based on existing exact solutions of the Gross-Pitaevskii
equation. Kagan {\it et al.} \cite{Kagan1997b} and Castin and Dum
\cite{Castin1996a} developed an analytical solution of the
time-dependent Gross-Pitaevskii equation for anisotropic
time-dependent harmonic potentials in the Thomas-Fermi
approximation. The time evolution of the Thomas-Fermi radii is
thereby described by three ordinary nonlinear differential
equations. To apply this solution we take advantage of the
separation of the center-of-mass motion, following the lines of
Japha and Band \cite{Japha2002a}. After deducing the equations of
motion we discuss the validity of this approach. We then apply the
theory to an anharmonic waveguide as it was studied in the
experiment. We conclude by introducing the corresponding
Hamiltonian formalism.

We start from the time-dependent Gross-Pitaevskii equation (GPE)
\begin{equation}
\label{gpe} i\hbar\frac{\partial}{\partial
t}\psi(\mathbf{x},t)=\left[-\frac{\hbar^2}{2m}\nabla^2+U(\mathbf{x})+g|\psi(\mathbf{x},t)|^2\right]\psi(\mathbf{x},t).
\end{equation}
We introduce a vector $\mathbf{R}(t)$ for an arbitrary
displacement of the condensate wave function and look for
solutions of the form
\begin{equation}
\label{ansatz}
\psi(\mathbf{x},t)=\psi_0(\mathbf{x}-\mathbf{R},t)e^{iS(\mathbf{x})}e^{i\beta(t)}.
\end{equation}
$\beta(t)$ denotes an additional global phase factor and the
gradient of the phase is given by the velocity
$\dot{\mathbf{R}}(t)$ of the displaced wave function
\begin{equation}
\label{spatialphase}
\nabla S(\mathbf{x},t)=\frac{m}{\hbar}\dot{\mathbf{R}}.
\end{equation}
Without loss of generality, the potential can be expressed in a
Taylor series about $\mathbf{R}$
\begin{eqnarray}
U(\mathbf{x})&=&U(\mathbf{R})+(\mathbf{x'}\nabla)
U(\mathbf{R})+\frac{1}{2}(\mathbf{x'}\nabla)^2U(\mathbf{R})+
...\nonumber\\
&=&\sum_{n=0}^{\infty}\frac{(\mathbf{x'}\nabla)^n}{n!}U(\mathbf{R})\label{potexpand}
\end{eqnarray}
In equation (\ref{potexpand}) we have used the abbreviation
\begin{equation}
(\mathbf{x'}\nabla)^n=\sum_{i_1,i_2,...,i_n=1}^3x'_{i_1}x'_{i_2}...x'_{i_n}\partial_{i_1}\partial_{i_2}...\partial_{i_n}.
\end{equation}
The shifted coordinates are given by
\begin{equation}
\label{derivative} x'_i=x_i-R_i
\end{equation}
and the spatial derivatives are identical in both coordinate
systems
\begin{equation}
\frac{\partial}{\partial x'_i}=\frac{\partial}{\partial x_i}.
\end{equation}
Inserting the wave function (\ref{ansatz}) and the potential
(\ref{potexpand}) into the Gross-Pitaevskii equation (\ref{gpe}),
we separate the result into three equations. The global phase
$\beta(t)$ is given by
\begin{equation}
\beta(t)=\frac{m}{\hbar}\int_0^t
dt'\left[-\ddot{\mathbf{R}}(t')\mathbf{R}(t')-\frac{1}{2}\dot{\mathbf{R}}(t')^2-\frac{1}{m}
U(\mathbf{R}(t'))\right].
\end{equation}
For the displacement vector $\mathbf{R}$ we find
\begin{equation}\label{eq1}
m\ddot{\mathbf{R}}=-\nabla U(\mathbf{R}).
\end{equation}
Summing up all others terms leads to
\begin{eqnarray}\label{eq2} i\hbar\frac{\partial}{\partial
t}\psi_0(\mathbf{x'},t)=&-&\frac{\hbar^2}{2m}\nabla^2\psi_0(\mathbf{x'},t)+g|\psi_0(\mathbf{x'},t)|^2\psi_0(\mathbf{x'},t)\nonumber
\\
&+&\left[\sum_{n=2}^{\infty}\frac{(\mathbf{x'}\nabla)^n}{n!}U(\mathbf{R})\right]\psi_0(\mathbf{x'},t).
\end{eqnarray}
Note, that the sum in equation (\ref{eq2}) starts from $n=2$ and
 the equation is written in the displaced coordinates $x'_i$.

We now assume that higher order contributions in the sum of
equation (\ref{eq2}) are small compared to the quadratic term,
which means that the potential can be considered as locally
quadratic. Then, taking only the leading term in the sum of
equation (\ref{eq2}), one recognizes the GPE in the rest frame of
the condensate
\begin{eqnarray}
\label{gpe2} i\hbar\frac{\partial}{\partial
t}\psi_0(\mathbf{x'},t)=\left[-\frac{\hbar^2}{2m}\nabla^2+\sum_{i=1}^3\frac{1}{2}m\omega_i^2(t)x'^2_i+g|\psi_0(\mathbf{x'},t)|^2\right]\psi_0(\mathbf{x'},t).\nonumber\\
\end{eqnarray}
The time-dependent harmonic potential is given by
\begin{equation}\label{eq3}
\frac{1}{2}m\omega_i^2(t)x'^2_i=\frac{1}{2}{x'}^2_i\frac{\partial^2}{\partial^2x'_i}U(\mathbf{R}),\,\,\,\,\,\,\,i=1,2,3,
\end{equation}
provided that the potential separates in the three coordinates and
the mixed derivatives vanish. The solution of equation
(\ref{gpe2}) within the Thomas-Fermi approximation is given in
Refs. \cite{Kagan1997b,Castin1996a}. It consists of three scaling
equations for the Thomas-Fermi radii of the condensate:
\begin{equation}\label{eq4}
\ddot{\lambda}_i=\frac{\omega_{i0}^2}{\lambda_i\lambda_1\lambda_2\lambda_3}-\omega_i^2(t)\lambda_i,\,\,\,\,\,\,\,i=1,2,3.
\end{equation}
The $\lambda_i$ are scaling factors for the condensate radii
$r_i(t)=\lambda_ir_{i0}$ and $r_{i0}$ are the Thomas-Fermi radii
at $t=0$. The initial conditions are $\lambda_i(0)=1$ and
$\dot{\lambda}_i(0)=0$. The solution in the rest frame is given by
\cite{Castin1996a}
\begin{eqnarray}
\label{castinwellenfunktion}
\psi_0(\mathbf{x'},t)=&&\sqrt{\frac{\mu}{g}}\frac{1}{\sqrt{\lambda_1
\lambda_2\lambda_3}}\left(1-\sum_{i=1}^3\left(\frac{{x'}_i}{r_{i0}\lambda_i}\right)^2\right)^{1/2}
\\
&&\times
\exp\left[i\frac{m}{\hbar}\sum_{i=1}^3\frac{\dot{\lambda}_i
{x'_i}^2}{2\lambda_i}-i\frac{\mu}{\hbar}\int_0^{t}\frac{dt'
}{\lambda_1\lambda_2\lambda_3}\right].\nonumber
\end{eqnarray}
Equation (\ref{eq4}) is also valid for complex oscillation
frequencies which corresponds to negative curvatures
\cite{Castin2002a}. Thus, the dynamics of the condensate are
completely determined by the set of ordinary differential
equations (\ref{eq1}), (\ref{eq3}) and (\ref{eq4}). To get the
complete condensate wave function in the laboratory frame one has
to insert the shifted coordinates (\ref{derivative}) into
(\ref{castinwellenfunktion}). Together with (\ref{ansatz}) and
(\ref{spatialphase}) this yields
\begin{eqnarray}
\label{cwf}
\psi(\mathbf{x},t)=&&\sqrt{\frac{\mu}{g}}\frac{1}{\sqrt{\lambda_1
\lambda_2\lambda_3}}\left(1-\sum_{i=1}^3\left(\frac{x_i-R_i}{r_{i0}\lambda_i}\right)^2\right)^{1/2}
\nonumber\\
&&\times
\exp\left[i\frac{m}{\hbar}\sum_{i=1}^3\left(\frac{1}{2}\frac{\dot{\lambda}_ix_i^2}{\lambda_i}-\frac{\dot{\lambda}_iR_ix_i}{\lambda_i}+\dot{R}_ix_i\right)\right]
\nonumber\\
&&\times \exp\left[i\beta(t)-i\frac{\mu}{\hbar}\int_0^{t}\frac{dt'
}{\lambda_1\lambda_2\lambda_3}+\frac{m}{2\hbar}\frac{R_i^2\dot{\lambda}_i}{\lambda_i}\right].
\end{eqnarray}
The first line of (\ref{cwf}) is the square root of the density
distribution of the condensate. As it is symmetric about
$\mathbf{x}=\mathbf{R}$, the vector $\mathbf{R}$ coincides with
the center of mass and therefore equation (\ref{eq1}) describes
the motion of the center of mass. The second line of (\ref{cwf})
contains the spatially varying phase factors, arising from the
internal dynamics (first two terms) and the center of mass motion
(third term). Their gradient defines the velocity field of the
condensate. In the last line of (\ref{cwf}), all global phase
factors are collected.

We discuss briefly the consequences of the theoretical model. For
a parabolic, linear or vanishing potential, the right hand side of
expression (\ref{eq3}) is constant or zero and the solution
(\ref{cwf}) is exact within the Thomas-Fermi approximation. The CM
motion is completely decoupled from the internal degrees of
freedom. For anharmonic potentials, equation (\ref{eq3})
establishes the coupling between the external and internal
dynamics. Equation (\ref{gpe2}) was derived under the assumption
that terms of third and higher order can be neglected in the sum
of equation (\ref{eq2}). This is a valid approximation as long as
the potential curvature varies only slightly over the extension of
the condensate. If the condition
\begin{equation}\label{cond}
\frac{\partial^3}{\partial
x^3}U(\mathbf{R})r_{0}\ll\frac{\partial^2}{\partial
x^2}U(\mathbf{R})
\end{equation}
is fulfilled the potential can be considered as locally quadratic.
The inequality (\ref{cond}) is violated when the quadratic part of
the potential is zero or becomes negative. In this case the
approximation remains valid if the contributions of the potential
energy in (\ref{eq2}) to the total energy are much smaller than
the chemical potential and the kinetic energy.

We now apply the theory to an anharmonic waveguide. This geometry
is the object of an actual study of the nonlinear condensate
dynamics \cite{Ott2002c}. The potential of the waveguide is
anharmonic in the $z$--direction and cylindrically symmetric in
the $\rho$--direction:
\begin{equation}
\label{waveguidepot} U(z,\rho)=\frac{1}{2}m\omega_{\rho}^2\rho ^2
+az^2+bz^3+cz^4,
\end{equation}
where we have expanded the potential up to the fourth order. With
the initial conditions $R_\rho(0)=0$, $R_z(0)=R_0$ and
$\dot{\mathbf{R}}(0)=0$ the solution of (\ref{eq1}) can be written
as a Fourier series
\begin{eqnarray}
\label{rzloesung} R_z(t)&=&\sum_{n=1}^\infty
a_n\cos{(n\omega_0 t)} \\
R_\rho(t)&=&0
\end{eqnarray}
$\omega_0$ denotes the fundamental frequency, which depends on the
amplitude of the center-of-mass oscillation. In the case of {\it
time of flight} measurements (duration $t$), the amplitudes and
phases of the Fourier components in the series (\ref{rzloesung})
change:
\begin{eqnarray}
a_n'&=&a_n\sqrt{1+\omega_n^2t^2} \\
\phi_n'&=&\phi_n+\arctan\omega_nt.
\end{eqnarray}
Inserting (\ref{rzloesung}) and (\ref{waveguidepot}) into the
coupling equation (\ref{eq3}) one obtains
\begin{eqnarray}
\omega_z(t)^2&=&\frac{1}{m}\left(2a+6bR_z(t)+12cR_z^2(t)\right) \nonumber\\
&=&\frac{1}{m}\left[2a+6b\sum_{n=1}^\infty a_n\cos{(n\omega_0
t)}+12c\left(\sum_{n=1}^\infty a_n\cos{(n\omega_0
t)}\right)^2\right]\nonumber\\
 &\simeq&\frac{1}{m}\left(2a+6ca_1+6ba_1\cos(\omega_0t)+6ca_1^2\cos(2\omega_0t)\right).
\end{eqnarray}
In the second step we have made the approximation, that $a_1\gg
a_{n}$ for $n>1$, which means, that the dominant Fourier component
is at the fundamental frequency $\omega_0$. In the rest frame, the
condensate is driven in its axial direction with $\omega_0$. The
driving amplitude is directly proportional to the amplitude of the
center-of-mass motion $a_1$. Due to the contribution of the fourth
order in the potential (\ref{waveguidepot}), the condensate is
additionally driven at twice the fundamental frequency with a
quadratic dependence on $a_1$. Beside these two frequencies, the
(m=0) low-lying quadrupole resonance is offresonantly excited
\cite{Ott2002c}. For small amplitudes, the oscillation frequency
in the rest frame becomes constant and equal to the frequency of
the center-of-mass motion ($\omega_z=\omega_0=\sqrt{2a/m}$). For
strong excitation, the nonlinearity of the equations of motion for
the internal condensates dynamics (\ref{eq4}) produces a nonlinear
coupling of the collective excitations. A comparison with actual
experimental data shows an excellent agreement between theory and
experiment \cite{Ott2002c}.

Kagan {\it et al.} \cite{Kagan1997b} pointed out, that the
equations of motion for the internal condensate dynamics
(\ref{eq4}) can also be derived from a classical Hamiltonian. For
the free motion in a trapping potential the total energy of the
condensate is conserved and can be calculated for the wave
function (\ref{cwf})
\begin{eqnarray}\label{hamilton}
H&=&\langle E_{\rm kin}\rangle+\langle E_{\rm pot}\rangle+\langle
E_{\rm int}\rangle \nonumber\\
&=&\frac{1}{7}\mu\sum_i\frac{\dot{\lambda}_i^2}{\omega_{i0}^2}+\frac{1}{7}\mu\sum_i\frac{\partial^2
U(\mathbf{R}(t))}{\partial
x_i^2}\frac{\lambda_i^2}{m\omega_{i0}^2}+\frac{2}{7}\mu\frac{1}{\lambda_1\lambda_2\lambda_3}\label{eq5}\nonumber\\
&&+\frac{1}{2}m\sum_i\dot{R}_i^2+U(\mathbf{R}),\label{eq6}.
\end{eqnarray}
The first three terms in (\ref{eq5}) are the contributions from
the internal dynamics and the last two terms that of the
center-of-mass motion. Taking equation (\ref{hamilton}) as a
classical Hamiltonian with the canonical variables
\begin{equation}
q_{i}\equiv\{R_1,R_2,R_3,\lambda_1,\lambda_2,\lambda_3\}
\end{equation}
and the conjugated momenta
\begin{equation}
p_{i}\equiv\{m\dot{R}_1,m\dot{R}_2,m\dot{R}_3,\frac{2\mu}{7\omega^2_{10}}\dot{\lambda}_1,\frac{2\mu}{7\omega^2_{20}}\dot{\lambda}_2,\frac{2\mu}{7\omega^2_{30}}\dot{\lambda}_3\}.
\end{equation}
one can evaluate the Hamilton equations of motion
\begin{eqnarray*}
 \dot{q}_i&=&\frac{\partial H}{\partial p_i} \\
 \dot{p}_i&=&-\frac{\partial H}{\partial q_i}.
\end{eqnarray*}
We obtain the equations (\ref{eq4}) and the definition of the
time-varying oscillation frequencies (\ref{eq3}). For the
center-of-mass motion we find a slightly modified equation of
motion
\begin{equation}\label{eq8}
m\ddot{R}_i=-\frac{\partial}{\partial x_i
}U(\mathbf{R})-\frac{1}{14}r_{i0}^2\lambda_i^2\frac{\partial^3}{\partial
x_i^3}U(\mathbf{R}).
\end{equation}
The second term in (\ref{eq8}) is a small correction to
(\ref{eq1}) which accounts for the influences of the internal
dynamics to the center-of-mass motion. The correction is
proportional to the square of the condensate radii, whereas the
internal dynamics (\ref{eq4}) are independent of the extension of
the condensate. The modified equation of motion (\ref{eq8}) is
useful to test the numerical stability of the calculation although
its influence on the amplitude of the center-of-mass oscillation
is negligible.

In conclusion, we have presented an approximate solution to the
time-dependent Gross-Pitaevskii equation for anharmonic
potentials. We have shown, that the internal dynamics of the
condensate are coupled to the external center-of-mass motion. For
an anharmonic waveguide, we have analyzed the internal dynamics in
detail and good agreement with the experimental data is found
\cite{Ott2002c}. We have also introduced a classical Hamiltonian
from which the equations of motions can be derived and have
discussed the validity of the approximations.

We gratefully acknowledge financial support from the Deutsche
Forschungsgemeinschaft under Grant No. Zi/419-5.

\end{document}